\def\den{\hbox{den}}
\def\hc{\hbox{h.c.}}
\def\tr{\hbox{tr}}
\def\ln{\ell{n}}
\documentstyle[12pt]{article}

 {\parindent 0.5cm \textwidth 15.0cm \textheight
23.5cm \hoffset=-0.8cm \voffset=-3cm
  \let\LARGE=\large
 \let\large=\normalsize
\begin{document}
\begin{titlepage} \vspace{0.2in} \begin{flushright}
MITH-93/21 \\ \end{flushright} \vspace*{1.5cm}
\begin{center} {\LARGE \bf  The Spontaneous Breaking of Chiral Symmetry without
Goldstone Bosons
\\} \vspace*{0.8cm}
{\bf She-Sheng Xue$^{a)}$}\\ \vspace*{1cm}
INFN - Section of Milan, Via Celoria 16, Milan, Italy\\ \vspace*{1.8cm}
{\bf   Abstract  \\ } \end{center} \indent

Considering a self-interaction only of mirror fermions in the context of a
lattice-regularized fermion field theory, we show that the system undergoes
spontaneous breaking of chiral symmetry and mirror-fermion masses are
generated. However, it is explicitly shown that there are no Goldstone bosons
appearing together with this spontaneous symmetry breaking phenomenon, since
Lorentz invariance, one of very general prerequisites of the Goldstone
theorem, is violated. The result and its possible application are briefly
discussed.
\vfill \begin{flushleft} 10th January, 1994\\
PACS 11.15Ha, 11.30.Rd, 11.30.Qc  \vspace*{3cm} \\
\noindent{\rule[-.3cm]{5cm}{.02cm}} \\
\vspace*{0.2cm} \hspace*{0.5cm} ${}^{a)}$
E-mail address: xue@milano.infn.it\end{flushleft} \end{titlepage}

\noindent
{\bf 1.}\hspace*{0.3cm} The ``Goldstone theorem'' \cite{goldstone,nambu} has
played a very important role in modern physics, in particular,
theoretical particle physics. The theorem states: ``there appear spinless
particles of zero mass whenever a continuous symmetry group leaves the
Lagrangian but not the vacuum invariant \cite{gold}.'' Its application to the
Standard
Model (SM), i.e., the spontaneous symmetry breaking phenomenon \cite{higgs},
succeed in cooperating the gauge symmetry $SU_L(2)\otimes U_Y(1)$ of the (SM)
with the massive gauge bosons $W^\pm, Z^\circ$ observed phenomenologically.
Three Goldstone bosons become the longitudinal modes of the massive gauge
bosons. However, we have never observed any extra massless modes of
Goldstone type in the elementary spectrum although this spectrum shows a total
violation of the chiral (flavour) symmetry of the SM. Thus, possibilities of
violating the chiral symmetry of the SM without
Goldstone modes deserve study. In order to avoid the appearance of extra
massless modes, one
might consider an explicit breaking of chiral symmetries through either
the fermionic sector \cite{ep} or anomalies in the gauge sector \cite{th}. In
this
note, we attempt to showing the possibility that a lattice-regularized fermion
field theory, which is chirally symmetric, undergoes a spontaneous symmetry
breaking without Goldstone modes appearing. However, the absence of massless
bosons is a consequence of the inapplicability of the ``Goldstone theorem''
rather than a contradiction of it, since the spontaneous breaking of chiral
symmetry to be considered does not obey one of the very general prerequisites
of
the ``Goldstone theorem'', i.e., Lorentz invariance, which is a necessary
condition in demonstrating the theorem \cite{gold,lorentz}.

\vspace*{0.5cm}
\noindent
{\bf 2.}\hspace*{0.3cm}
We thus consider a chirally symmetric lattice-regularized fermion field theory.
There must exist ordinary and mirror fermions in the low-energy
spectrum of the theory \cite{nogo}. As for the Wilson fermion \cite{wilson}, a
dimension-5 operator (the Wilson term) in the continuum limit is introduced
($S_d$ is a massless Dirac action)
\begin{equation}
S=S_d(\bar\psi,\psi)+{r\over a}\sum_{x,\mu}\bar\psi(x)\partial_\mu^2\psi(x),
\label{wilson}
\end{equation}
where $a$ is the lattice spacing and the lattice ``laplacian'' $\partial_\mu^2$
is defined as
\begin{equation}
\partial_\mu^2\psi(x) = \psi(x+a_\mu)+\psi(x-a_\mu)-2\psi(x).
\label{laplacian}
\end{equation}
This high-dimension operator becomes an irrelevant operator in the infrared and
does not affect the continuum limit at all. However, chiral symmetry in
(\ref{wilson})
is explicitly broken. One might introduce a self-interaction of mirror
fermions only, and generalize (\ref{wilson}) to
\begin{equation}
S=S_d(\bar\psi,\psi)-{G_2\over 2}\sum_{x,\mu}
(\bar\psi_L(x)\partial_\mu^2\psi_R(x))(\bar\psi_R(x)\partial_\mu^2\psi_L(x)).
\label{action}
\end{equation}
Analogous to the Wilson term (\ref{wilson}), this dimension-10 operator in the
continuum limit is not relevant for ordinary fermions with $ka\sim 0$. However,
for mirror fermions with $ka\sim \pi$, this operator is not irrelevant and it
presents a self-interaction of these mirror fermions at the cutoff scale.
Obviously, eq.(\ref{action}) is chiral invariant. Note that in general, there
are several possibilities of chirally invariant four fermion interactions on a
lattice space time, e.g.,
\begin{eqnarray}
&&\beta_1\bar\psi(x)\psi(x)\bar\psi(x)\psi(x)+
\beta_1\sum_\mu\bar\psi(x\pm\mu)\psi(x)\bar\psi(x)\psi(x)\nonumber\\
&&+\beta_3\sum_{\mu\nu}\bar\psi(x\pm\mu)\psi(x\pm\nu)\bar\psi(x)\psi(x)
+\cdot\cdot\cdot,
\label{four}
\end{eqnarray}
which represent complicated interactions between ordinary fermions and mirror
fermions. The origin of these interactions (which might stem from the quantum
gravity \cite{planck,xue} or a high-frequency contribution of the theory
\cite{bardeen}) will not be a focus of this paper. We have tuned the
four-fermion couplings in eq.(\ref{four}) such that only mirror fermions
interact with themselves (\ref{action}) and the interactions between ordinary
fermions and mirror fermions vanish.

On the other hand, in order to generate effective masses $M$ of order
$O({1\over a})$ for mirror fermions, which split mirror fermion from ordinary
fermions as occurs in (\ref{wilson}), one expects that the Fermi-type coupling
$G_2$ be large enough so that theory (\ref{action}) undergoes spontaneous
symmetry breaking with non-vanishing vacuum expectation value ($r=\bar r a$)
\begin{equation}
\bar r={G_2\over 4}\!\sum_{\mu,x}\left({1\over 4}\right )
\left\langle\bar\psi_L(x)\partial_\mu^2\psi_R(x)+\hc\right\rangle
.\label{r}
\end{equation}
This belief can be justified based on computation of the effective
potential and gap-equation of the system by adopting the Landau mean-field
method and the large-$N_f$ approach ($G_2N_f$ fixed, $N_f \gg 1$ is the number
of flavour in eq.(\ref{action})). The effective potential is given by
($\int_l = \int^\pi _{- \pi} {d^4l\over (2\pi)^4}$)
\begin{equation}
V(\bar r)={4\bar r^2\over G_2}-N_f\tr\int_l\ln
\left[ {\gamma_\mu\sin l_\mu\over a}+\bar rw(l)\right]+\cdot\cdot\cdot,
\label{eff}
\end{equation}
where $l_\mu = p_\mu a; w(l)=\sum_\mu(1-\cos l_\mu)$,
the first term is the Landau mean-field approximations (tree-level),
the second term is a one-loop approximation (with mean-fields) in the
large-$N_f$
approach, both are $O(N_f)$ terms, and the
dots denote $O(N_f^0)$ terms. In eq.~(\ref{eff}) the imaginary part,
$i{G_2\over 4}\!\sum_{\mu,x}\left({1\over 4}\right )
\langle\bar\psi_L(x)\partial_\mu^2\psi_R(x)-\hc\rangle,$
can always be gauged away by
an appropriate transformation because of the chiral symmetry of the action
(\ref{action}). By minimizing the lowest-order effective potential
$V(\bar r)$ eq.~(\ref{eff}), we readily obtain a self-consistent
``gap equation'':
\begin{equation}
r = {g_2\over 2} \int_l w(l) { rw(l) \over \den(l)},
\label{gap2}
\end{equation}
where $g_2 a^2 = N_f G_2$ and $\den(l) = \sin^2l_\mu +(rw(l))^2$.
This gap equation is exact for $N_f\rightarrow\infty$.
Numerical calculation shows, for $g_2 > 0.2$ \footnote{The critical value
is not very large, approximation is good even for small $N_f$.}, there exist
a non-trivial solution $r > 0$. One clearly finds that the solution $r > 0$
stems from the contribution of mirror fermions, owing to the factor $w(l)$ in
eq.~(\ref{gap2}). The trivial solution $r=0$ cannot be realized since it
does not represent the true ground state ($\Delta E=V(r)-V(0)$)
\begin{equation}
\Delta E= -{2N_f \over a^4} \int_l \sum^{\infty} _{k=1}
{1 \over k+1}
\left[{(rw(l))^2 \over s^2(l) + (rw(l))^2} \right]^{k+1}+O(N_f^0),
\label{vac1}
\end{equation}
which is obtained by substituting (\ref{gap2}) into (\ref{eff}). Thus, the
system does undergo spontaneous symmetry breaking. Mirror-fermion masses
are generated through their self-interactions (Nambu-Jona Lasinio mechanism
\cite{nambu}). The effective action
above the
true ground state ($r\not=0$) is just the Wilson fermion (\ref{wilson}).

\vspace*{0.5cm}
\noindent
{\bf 3.}\hspace*{0.3cm}
Are there Goldstone modes appearing together with the spontaneous symmetry
breaking discussed above? We discuss this question by performing an explicit
calculation of the scattering amplitude of mirror fermions and anti
mirror-fermions.
The self-interaction of mirror fermions in action (\ref{action})
can be written as
\begin{equation}
-{G_2\over a}\sum_{x,\mu}\left({1\over 4}\right )\left[
(\bar\psi(x)\partial_\mu^2\psi(x))(\bar\psi(x)\partial_\mu^2\psi(x))-
(\bar\psi(x)\gamma_5\partial_\mu^2\psi(x))(\bar\psi(x)\gamma_5\partial_\mu^2
\psi(x))\right],
\label{action1}
\end{equation}
which clearly gives us the scattering vertices of mirror fermions in the scalar
channel ($s$) and the pseudo-scalar channel ($p$)  (see Fig.1)
\begin{eqnarray}
V_i&\sim& (-)^i{G_2\over2}\sum_{\mu\nu}\delta_{\mu\nu}T_{\mu\nu}(k,q),\hskip2cm
i=p,s\\
T_{\mu\nu}&=&(1-\cos(k+{q\over2})_\mu)(1-\cos(k+{q\over2})_\nu),
\label{vertices}
\end{eqnarray}
where $(-)^p=-1,\, (-)^s=1$ and the momentum $q$ for composite particles is
much smaller than
the momentum $k$ of mirror fermions $q\ll k\sim \pi$. Then, we consider a
one-loop
bubble diagram (see Fig.2)
\begin{equation}
\sum_{\mu\nu}\left((-)^i{G_2\over 2}\right
)^2T_{\mu\nu}(k,q)B^{(i)}_{\mu\nu}(q)
\label{bubble}
\end{equation}
with
\begin{equation}
B^{(i)}_{\mu\nu}(q)=-{1\over a^2}\int_l\tr\left[
\Gamma_iS(l+{q\over2})\Gamma_iS(l-{q\over2})\right]T_{\mu\nu}(l,q),
\label{bubble1}
\end{equation}
where $\Gamma_p=\gamma_5, \Gamma_s=1$ and the mirror-fermion propagator $S$
satisfying gap-equation (\ref{gap2}) is given by
\begin{equation}
S(l\pm{q\over2})={-\gamma_\mu s_\mu(l\pm {q\over2})+rw(l\pm {q\over2})\over
\den(l\pm{q\over 2})},
\label{propagator}
\end{equation}
where $s_\mu(l)=\sin l_\mu$. Straightforward calculation leads us to
\begin{equation}
B^{(i)}(q)=(-)^i{4N_f\over a^2}\int_l{s^2(l)-(-)^i(rw(l))^2\over [\den(l)]^2}
T_{\mu\nu}(l)+Q(q)E^{(i)}_{\mu\nu}+O[(q^2)^2],
\label{b}
\end{equation}
where $Q(q)={4\over a^2}\sum_\mu\sin^2{q_\mu\over 2}$ and
\begin{equation}
E^i_{\mu\nu}=(-)^i{N_f\over 4}\left[-\int_l{c^2(l)-(-)^ir^2s^2(l)\over
[\den(l)]^2}
T_{\mu\nu}(l)+\delta_{\mu\nu}\int_l{s^2(l)-(-)^i(rw(l))^2\over [\den(l)]^2}
s^2(l)\right].
\label{e}
\end{equation}
Note that throughout our calculation, the operators $\partial_\mu^2,\,w(l)$
and $T_{\mu\nu}(l)$ can be understood as the projective operators to pick out
mirror components of fermion field. Now we are in a position to consider the
chain approximation (leading order in
large-$N_f$ expansion) to the scattering amplitude of mirror fermions and anti
mirror fermions,
$A^{(i)}\sim\sum_\mu\langle\bar\psi(x)\partial_\mu^2\Gamma_i\psi(x)
\bar\psi(0)\partial_\mu^2\Gamma_i\psi(0)\rangle$ (see Fig.3)
\begin{eqnarray}
A_{(i)}(q^2)&=&(-)^i{G_2\over 2}\sum_{\mu\nu}\delta_{\mu\nu}T_{\mu\nu}(k,q)
+\left((-)^i{G_2\over 2}\right
)^2\sum_{\mu\nu}T_{\mu\nu}(k,q)B^{(i)}_{\mu\nu}(q)
+\cdot\cdot\cdot\\
&=&{G_2\over 2}\sum_{\mu\nu}T_{\mu\nu}(k,q)\left({1\over
\delta_{\mu\nu}-(-)^i{G_2\over 2}B^{(i)}_{\mu\nu}(q)}\right),
\label{a}
\end{eqnarray}
where
$(B^{(i)}_{\mu\nu}(q))^n=B^{(i)}_{\mu\sigma}(q)B^{(i)}_{\sigma\rho}(q)\cdot\cdot
\cdot B^{(i)}_{\lambda\nu}(q)$. Substituting the $\delta_{\mu\nu}$ in
denominator of (\ref{a}) by the gap-equation (\ref{gap2})
\begin{equation}
\delta_{\mu\nu}=\delta_{\mu\nu}{g_2\over 2} \int_l w(l) { w(l) \over \den(l)},
\end{equation}
and
$B^(i)_{\mu\nu}(q)$ by eq.~(\ref{b}), we examine the
poles of (\ref{a}). These poles are located at
\begin{equation}
{G_2N_f\over a^2}\left(\delta_{\mu\nu}{1\over 2}\int_l{(w(l))^2\over \den(l)}
-2\int_l{T_{\mu\nu}(l)\over \den(l)}\right ).
\label{pp}
\end{equation}
for the pseudo-scalar channel and
\begin{equation}
{G_2N_f\over a^2}\left(\delta_{\mu\nu}{1\over 2}\int_l{(w(l))^2\over \den(l)}
-2\int_l{s^2(l)-(rw(l))^2\over [\den(l)]^2}T_{\mu\nu}(l)\right ).
\label{ps}
\end{equation}
for the scalar channel. One finds that (\ref{pp}) does not vanish and
eq.~(\ref{a}) does not possess a zero pole. Thus, we do not find
any massless modes in the pseudo-scalar channel of the scattering amplitudes of
mirror fermions. Actually, the non-vanishing of eqs.~(\ref{pp}) and (\ref{ps})
tells us that the composite modes of mirror fermions have their masses at the
order
of the cutoff and are then unobservable in the low-energy spectrum. However,
instead of the self-interaction of mirror fermions introduced, we consider
four-fermion interaction $G_1\bar\psi(x)\psi(x)\bar\psi(x)\psi(x)$ with chiral
symmetry breaking v.e.v. $\langle\bar\psi(x)\psi(x)\rangle\not=0$. Then,
$\partial_\mu^2\sim
1$, $T_{\mu\nu}(l)\sim\delta_{\mu\nu}$ and $w(l)=\sum_{\mu\nu}T_{\mu\nu}
\sim\sum_{\mu\nu}\delta_{\mu\nu}$ are no longer projective operators for mirror
fermions, eq.~(\ref{pp}) vanishes and one obtains a Goldstone boson in the
pseudo-scalar channel. For the scalar channel, the composite scalar is still
massive at the order of the cutoff due to the contribution of mirror fermions.

\vspace*{0.5cm}
\noindent
{\bf 4.}\hspace*{0.3cm}
We turn to discuss the reasons for the absence of Goldstone modes. Lattice
regularization of a fermion field theory ($S_d(\bar\psi,\psi)$ in
(\ref{wilson})
) by itself is not explicitly Lorentz invariant, in particular, the
high-dimension operators in eqs.~(\ref{wilson},\ref{action}) are
Lorentz-symmetry violating. Lorentz symmetry is expected to be restored in
the continuum limit, where these high-dimension operators are irrelevant. If we
consider the scattering vertex
four-fermion interaction $G_1\bar\psi(x)\psi(x)\bar\psi(x)\psi(x)$
with chiral symmetry breaking
$\langle\bar\psi(x)\psi(x)\rangle\not=0$ and
$A^{(p)}\sim\sum_\mu\langle\bar\psi(x)\gamma_5\psi(x)
\bar\psi(0)\gamma_5\psi(0)\rangle$, all of them are Lorentz invariant and
remain relevant operators in the continuum limit, where Lorentz symmetry
is fully restored and we have Goldstone modes as those in the continuum
Nambu-Jona Lasinio model. However, in the context we discuss, the scattering
vertex (\ref{action}), chiral symmetry breaking (\ref{r}) and scattering
amplitude (\ref{a}) of mirror fermions, whose momenta probe the detail of
lattice structure, violate Lorentz invariance and these operators
turn out to be irrelevant in the continuum limit. Thus, the approach of
demonstrating the Goldstone theorem, where the Lorentz invariance of the theory
is a necessary prerequisite, cannot be applied. Although the self-coupling
$G_2$ of mirror fermions can be strong enough to break chiral symmetry,
a Goldstone mode is not bound to be produced.

Starting from the Goldstone theorem, let us suppose that there is a Goldstone
mode appearing, corresponding to the spontaneous breaking of chiral symmetry
$r\not=0$ (\ref{gap2}), which leads to the effective Wilson action
(\ref{wilson}) through asymmetric vacuum (\ref{vac1}). However, the
chiral-symmetry breaking term, i.e., Wilson term (\ref{wilson}), turns out to
be irrelevant and chiral symmetry and Lorentz symmetry are restored as the
continuum limit is approached, the Goldstone boson, which is supposed to be
produced, would not disappear in this limit owing to its long-range property.
Then, in the continuum limit, we would have obtained the Goldstone boson
without any symmetry breaking. This turns out to be in contradiction with the
Goldstone theorem itself, which should be true in the continuum limit.
Thus, this logical argument leads us to the absence of the Goldstone modes in
the spontaneous symmetry breaking phenomenon we discussed, although we have not
presented a rigorous demonstration.

Once the Wilson term is generated without Goldstone bosons, nothing can prevent
the dimension-3 operator $m\bar\psi(x)\psi(x)$ \cite{bardeen,xue1} and the
anomaly \cite{smit} from being produced if we turn on gauge interactions. Also,
we might find a way out of the problem of the non-conservation of fermion
numbers \cite{th} in a lattice-regularized version of the Standard Model.

\newpage  \pagestyle{empty}
\begin{center} \section*{Figure Captions} \end{center}

\vspace*{1cm}

\noindent {\bf Figure 1}: \hspace*{0.5cm}
The scattering vertices of mirror fermions.

\noindent {\bf Figure 2}: \hspace*{0.5cm}
One-loop bubble diagram.

\noindent {\bf Figure 3}: \hspace*{0.5cm}
The scattering amplitude of mirror fermions in the chain approximation.

}

\begin{thebibliography}{99}

\bibitem{goldstone}
J.~Goldstone, {\sl Nuovo Cimento} {\bf 19} (1961) 154.

\bibitem{nambu}
Y.~Nambu and G.~Jona-Lasinio, {\sl Phys. Rev.} {\bf 122} (1961) 345.

\bibitem{gold}
J.~Goldstone, A.~Salam and S.~Weinberg, {\sl Phys.~Rev.} {\bf 127} (1962) 965.

\bibitem{higgs}
P.~W.~Higgs, {\sl Phys.~Lett.} {\bf 12} (1964) 132, {\sl Phys.~Rev.}
{\bf 145} (1966) 1156;\\
F.~Englert and R.~Brout, {\sl Phys.~Rev.~Lett.} {\bf 13} (1964) 321.

\bibitem{ep}
E.~Eichten and J.~Preskill, {\sl Nucl.~ Phys.} {\bf B268} (1986) 179.

\bibitem{th}
G.~'t Hooft, {\sl Phys.~Rev.~Lett.} {\bf 37} (1976) 8;
{\sl Phys.~Rev.} {\bf D14} (1976) 3432.

\bibitem{lorentz}
A.~Klein and B.W.~Lee, {\sl Phys.~Rev.~Lett.} {\bf 12} (1964) 266;\\
W.~Gilbert, {\sl Phys.~Rev.~Lett.} {\bf 12} (1964) 713;\\
G.S.~Guralnik, C.R.Hagen and T.W.B.~Kibble, {\sl Phys.~Rev.~Lett.} {\bf 13}
(1964) 585.

\bibitem{nogo}
H.B.~Nielsen and M.~Ninomiya, {\sl Nucl.~Phys.} {\bf B185} (1981) 20, {\it
ibid.} {\bf B193} (1981) 173, {\sl Phys.~Lett.} {\bf B105} (1981) 219.

\bibitem{wilson}
K.~Wilson, in {\it New phenomena in subnuclear physics\/}
(Erice, 1975)
ed.\ A.~Zichichi (Plenum, New York, 1977).

\bibitem{planck}
C.W.~Misner, K.S.~Thorne and J.A.~Wheeler, {\it Gravitation\/} (Freeman,
San Fransisco, 1973).

\bibitem{xue}
G.~Preparata and S.-S.~Xue, {\sl Phys.~Lett.} {\bf B264} (1991) 35;
{\sl Nucl.~Phys.} {\bf B26} (Proc.~Suppl.) (1992) 501;
{\sl Nucl.~Phys.} {\bf B30} (Proc.~Suppl.) (1993) 647.

\bibitem{bardeen}
W.A.~Bardeen, C.T.~Hill and M.~Linder,
{\sl Phys. Rev. Lett.} {\bf 56} (1986) 1230,
{\sl Nucl. Phys.} {\bf B273} (1986) 649 and {\it ibid.} {\bf B323} (1989)
493,\\
W.A.~Bardeen, C.~T.~Hill and M.~Lindner {\sl Phys.\ Rev.} {\bf D41} (1990)
1647,
\\
W.A.Bardeen, S.~T.~Love and V.A.~Miransky, {\sl Phys.\ Rev.} {\bf D42} (1990)
3514,\\
A.~Koci\'c, S.~Hands, J.B.~Kogut and E.~Dagotto, {\sl Nucl. Phys.} {\bf B347}
(1990) 217.

\bibitem{xue1}
G.~Preparata and S.-S.~Xue, {\sl Phys.~Lett.} {\bf B302} (1993) 442;
S.-S.~Xue, ``Aspects of dynamical symmetry breaking on a lattice'', MITH 93/20.

\bibitem{smit} L. H. Karsten and J. Smit, {\sl Nucl. Phys.} {\bf B144}
(1978) 536.

\end{thebibliography}
\end{document}